\documentstyle[prl,aps,preprint]{revtex}
\begin{document}
\draft
\title{Long wavelength instability for uniform shear flow}
\author{Mirim Lee and James W. Dufty}
\address{Department of Physics, University of Florida \\
Gainesville, FL 32611}
\author{Jos\'e M. Montanero and Andr\'es Santos}
\address{Departamento de F\'{\i}sica, Universidad de Extremadura, \\
E-06071 Badajoz, Spain}
\author{James F. Lutsko}
\address{ESADG, Dept. of Chem. Eng., Katholiek University of Leuvan\\
B-3001, Heverlee, Belgium}
\date{\today }
\maketitle

\begin{abstract}
Uniform shear flow is a prototype non-equilibrium state admitting detailed
study at both the macroscopic and microscopic levels via theory and computer
simulation. It is shown that the hydrodynamic equations for this state have
a long wavelength instability. This result is obtained first from the
Navier-Stokes equations and shown to apply at both low and high densities.
Next, higher order rheological effects are included using a model kinetic
theory. The results are compared favorably to those from Monte Carlo
simulation.
\end{abstract}

\pacs{PACS numbers: 47.20.Ft, 05.20.Dd, 05.60.+w, 0.5.70Ln}

%\narrowtext

\narrowtext
In spite of extensive formal theory for non-equilibrium statistical
mechanics, definitive tests and controlled illustrations outside the domain
of linear response are rare. One such rare case is the macroscopic state of 
{\em uniform} shear flow. Like planar Couette flow the $x$-component of the
average flow velocity varies linearly in the $y$ direction. However, the
flow is generated by periodic boundary conditions in the local Lagrangian
frame \cite{edwards,dufty1,dufty2} leading to a uniform temperature and
pressure, with a monotonic increase of the temperature. This is in contrast
to planar Couette flow, driven by local boundary conditions, for which the
temperature is non-uniform but stationary \cite{BSD}. The advantage of
uniform shear flow is that the boundary conditions allow application of
computer simulation methods with periodic imaging to emulate bulk effects
for small systems, just as in the simulation of equilibrium states. In
addition, there are many theoretical advantages such as the absence of a
hydrodynamic boundary layer. The problems associated with increasing
temperature can be controlled by the addition of a thermostat so that a
steady state with the desired flow field results. During the past fifteen
years uniform shear flow has been the focus of attention in most studies of
classical non-equilibrium statistical mechanics by a wide range of
theoretical and simulation methods \cite{hanley,mareschal}. It is one of the
few means by which rheology and transport far from equilibrium can be
studied at the fundamental level. More recently, simulations of uniform
shear flow have provided the test for new concepts of non-equilibrium
variational principles \cite{evans1} and for the relationship between
transport and chaotic Hamiltonian flows\cite{evans2}.

Although it has been known for more than ten years that uniform shear flow
undergoes a transition to an ordered state at large shear rates \cite
{erpenbeck}, its stability at smaller shear rates has not been questioned.
Our objective here is to report an instability of uniform shear flow at any
shear rate for a sufficiently long wavelength perturbation. More precisely,
solutions to the hydrodynamic equations linearized about the macroscopic
state of uniform shear flow show exponential growth in time for wavenumbers
smaller than a critical wavenumber $k_{\text{cr}}(a) > 0$ for
$a > 0$. For the case of zero shear rate these linear equations define the
hydrodynamic modes which describe how small perturbations of the velocity,
temperature, and density fields decay back to equilibrium. These are the two
sound modes, a heat mode, and two shear modes. In a similar way the
hydrodynamic modes describing response to 
perturbations of any reference
stationary state can be identified. If all such modes decay in time the
reference stationary state is referred to as linearly stable; conversely, if
one or more modes leads to an increasing amplitude in time the reference
state is linearly unstable. 

The hydrodynamic equations are approximations to the more general local
conservation laws for the mass, energy, and momentum densities, 
\begin{equation}
D_tn+n\nabla \cdot {\bf U}=0,  \label{1}
\end{equation}

\begin{equation}  \label{2}
D_t e + (e + p)\nabla \cdot {\bf U} + \nabla \cdot{\bf S} +
P_{ij}\partial_iU_j = w,
\end{equation}

\begin{equation}
D_tU_i+\rho ^{-1}\partial _ip+\rho ^{-1}\partial _jP_{ij}=0,  \label{3}
\end{equation}
where $D_t\equiv \partial _t+{\bf U}\cdot \nabla $ is the material
derivative. The momentum density is related to the flow field ${\bf U}$ by $%
{\bf g}=\rho {\bf U}$, $\rho =mn$ is the mass density ($m$ = mass, 
$n$ = number
density), and $e$ is the internal energy density. The pressure, $p$, is not
an independent variable but is {\em defined} to have the same functional
relationship to $n$ and $e$ as in equilibrium. The inhomogeneous term on the
right side of the energy equation is due to an external non-conservative
force representing the thermostat. There are several thermostats that have
been used in simulations and theory. Here the thermostat is fixed by a force
on each particle proportional to the velocity relative to the local flow
field. Consequently, its average value is zero and the effect of this force
occurs only in the energy equation. The proportionality constant is
determined by requiring stationarity for the uniform shear flow state.
The resulting term $w$ in Eq. (2) is the same as that obtained
using the thermostat for simulations.
Finally, the irreversible heat and momentum fluxes are denoted by ${\bf S}$
and $P_{ij}$, respectively. These equations are exact but incomplete until
the irreversible fluxes are specified in terms of the hydrodynamic fields.
Nevertheless, it is possible to define the state of uniform shear flow
without their detailed form. Steady state uniform shear flow is defined by
constant (in both space and time) energy and mass densities, $\{e_0,\rho _0\}
$, and a flow velocity whose only non-vanishing component is $U_{0x}=ay$.
The constant, $a$, is the shear rate and provides the single control
parameter measuring the deviation from equilibrium. The boundary conditions
are simple periodic conditions in the local Lagrangian coordinate frame, $%
{\bf r^{\prime }}={\bf r}-{\bf U}_0({\bf r})t$. Substitution of these
assumptions for $e,\rho ,$ and ${\bf U}$ into the above conservation laws
shows they are all satisfied if ${\bf S}$ and $P_{ij}$ are also uniform and
the parameter of the thermostat in the energy equation is chosen such that $%
w_0=aP_{0xy}$.

Consider small deviations of the hydrodynamic variables from the uniform
shear flow state. To be explicit it is necessary to specify the heat and
momentum fluxes. We first consider the case for which all spatial gradients
are small, including the shear rate $a$. Then the heat flux is given by
Fourier's law and the momentum flux is given by Newton's viscosity law, 
\begin{equation}
{\bf S}=-\lambda \nabla T,  \label{4}
\end{equation}

\begin{equation}
P_{ij}=-\eta (\partial _iU_j+\partial _jU_i-\case{2}{3}\nabla \cdot {\bf U}%
\delta _{ij})-\kappa \nabla \cdot {\bf U}\delta _{ij},  \label{5}
\end{equation}
where $\lambda $, $\eta $, and $\kappa $ are the thermal conductivity, shear
viscosity, and bulk viscosity, respectively. These are the leading terms in
a ``uniformity'' expansion ordered according to spatial gradients of the
hydrodynamic fields \cite{McL}. It is convenient to choose density and
temperature as independent thermodynamic parameters and to denote the
deviations in the hydrodynamic parameters from their values for uniform
shear flow by $y_\alpha \equiv \{\delta n,\delta T,\delta U_i\}$. Also, we
consider a fluid of hard spheres to specify the equations of state, $e=e(n,T)
$ and $p=p(n,T)$. These are determined from the Percus-Yevick equation \cite
{hansen} which is known to be accurate over the entire fluid phase.
Combining Eqs.~(\ref{1})--(\ref{5})
 and retaining terms only up through
linear order in $y_\alpha $ identifies the hydrodynamic equations for small
perturbations about uniform shear flow. The boundary conditions are made
explicit by looking for solutions of the form 
\begin{equation}
y_\alpha ({\bf r},t)=\widetilde{y}_\alpha ({\bf k},t)\exp (i{\bf k}\cdot 
{\bf r^{\prime }}),  \label{6}
\end{equation}
where $\widetilde{y}_\alpha ({\bf k},t)$ is the amplitude of a mode with
wavelength, $2\pi /k$. To simplify the analysis and to focus on the
instability the following is restricted to the special case $k_z=k_x=0$,
i.e. spatial perturbations only along the velocity gradient. The linear
equations for $\widetilde{y}_\alpha ({\bf k},t)$ are then of the form 
\begin{equation}
\partial _t\widetilde{y}_\alpha ({\bf k},t)+M_{\alpha \beta }(a,{\bf k})%
\widetilde{y}_\beta ({\bf k},t)=0  \label{7}
\end{equation}
with \narrowtext
\begin{equation}
{\sf M}(a,{\bf k})=\left( 
\begin{array}{ccccc}
0 & 0 & 0 & nik & 0 \\ 
c_1a^2 & (\eta /2e)a^2+Dk^2 & -(2\eta T/e)aik & (pT/e)ik & 0 \\ 
-\nu _naik & -(\nu /2T)aik & \nu k^2 & a & 0 \\ 
(p_n/\rho )ik & (p/\rho T)ik & 0 & \gamma k^2 & 0 \\ 
0 & 0 & 0 & 0 & \nu k^2
\end{array}
\right) ,  \label{7bis}
\end{equation}
%\narrowtext
where $c_1\equiv -\rho \nu _nT/e$, $D\equiv \lambda T/e$, $\nu \equiv \eta
/\rho $, $\gamma \equiv \frac 43\nu +\kappa /\rho $, and $z_n\equiv \partial
z/\partial n$.

The dispersion relations obtained from $\det (s{\sf I}-{\sf M})=0$ give five
hydrodynamic modes, $s_\alpha (a,{\bf k})$. If the real parts of one or more
modes become negative for some values of $k$ and $a$ then solutions to 
Eqs.~(\ref{7}) grow in time and the uniform shear flow state is linearly
unstable. Direct calculation shows this is the case for sufficiently small $k
$ for any fixed and finite $a$. The essential mechanism for the instability
follows from the linearization of the viscous heating term $P_{ij}\partial
_iU_j$ in the equations for $\delta T$ which is marginally controlled by the
linearization of the thermostat. The hydrodynamic modes for finite $a$ and
asymptotically small $k$ can be calculated explicitly to order $k^2$. Four
modes vanish as $k\rightarrow 0$ while one is finite, 
\begin{equation}
s_0\rightarrow \nu k^2\hspace{.1in},s_1\rightarrow (\eta /2e)a^2\hspace{.1in}%
,\hspace{.1in}s_2\rightarrow c_2k^2\hspace{.1in},  \label{8}
\end{equation}
\begin{equation}
s_3=s_4^{*}\rightarrow ick-(d_1-d_2a^2)(k/a)^2\hspace{.1in},  \label{9}
\end{equation}
where $\rho c^2=\left[ 2p(2+n\nu _n/\nu )+np_n\right] $, $\rho c^2c_2=n(3\nu
p_n-2p\nu _n)$, $(\rho \nu )^2d_1=p\left[ 2e(2\nu +n\nu _n)-\nu p\right] $,
and $2d_2=(3\nu +\gamma -c_2)$. These are positive constants depending only
on the density and temperature of the reference state. The first three modes
are stable whereas the complex conjugate pair $s_3$ and $s_4$ are unstable
for $a^2 < d_1/d_2$. {\em This is a primary observation of our work}:
within the limitations of the well-established Navier-Stokes equations
(e.g., small $k$ and small $a$) the equations are unstable at all $k$ for a
reference state with sufficiently small shear rate. Conversely, a similar
expansion of the eigenvalues in powers of $a$ at fixed $k$ shows the
instability at all shear rates for sufficiently small $k$. These asymptotic
results are confirmed by an exact evaluation of the eigenvalues from 
(\ref{7bis}),
 using the Percus-Yevick approximation for the virial equation of state 
\cite{hansen} and using the Enskog kinetic theory to specify transport
coefficients \cite{ferziger}. The results are shown in Figure \ref{fig1},
indicating the line in the $k$--$a$ plane separating stable (above) from
unstable (below) dynamics at three different densities ($n^{*}\equiv n\sigma
^3$ where $\sigma $ is the hard sphere diameter). The instability is
qualitatively the same for all densities.

The Navier-Stokes approximation for the heat and momentum fluxes requires $%
k\ll $ inverse mean free path and $a\ll $ inverse mean free time. Since the
reference state is generated by the shear it is possible that higher order
contributions in $a$ might remove the long wavelength instability. To
address this question, it would be desirable to derive the linear
hydrodynamic equations from the Boltzmann-Enskog kinetic equation without
this limitation to small $a$. However, no solution to this kinetic equation
is known, for either the reference uniform shear flow state or deviations
from it. Consequently, we consider the case of low density for which the
Boltzmann equation applies. While no solutions to this kinetic equation are
known either, it is well-established that closely related ``kinetic models''
provide practical and accurate representations of solutions to the Boltzmann
equation. Here, we chose the non-linear BGK kinetic model obtained by
replacing the Boltzmann collision operator with an average collision
frequency times the deviation of the distribution function from a local
equilibrium distribution. The parameters of this local equilibrium
distribution are chosen to enforce the exact conservation laws. As a
consequence, the Navier-Stokes hydrodynamic equations obtained from the BGK
model are the same as those from the Boltzmann equation -- only the values of
the transport coefficients differ. On this basis, we expect that
hydrodynamic equations outside the Navier-Stokes limit derived from the BGK
model should provide a faithful representation of those that would be
obtained from the Boltzmann equation. These expectations for the BGK model
have been confirmed in the case of shear flow by 
both analytical and numerical comparisons with
the Boltzmann equation \cite{santos}.
It is remarkable that the BGK equation can be solved exactly to
determine the reference state distribution function for
uniform shear flow, at arbitrary shear rate 
\cite{santos2}. Using this known result, the heat and momentum fluxes in the
conservation laws (\ref{1})--(\ref{3})
 can be determined to leading order in the
deviations of the hydrodynamic fields from uniform shear flow. The resulting
hydrodynamic equations linearized about the reference state are again valid
only to order $k^2$ but now there is no {\em a priori} restriction on the
value of the shear rate; the form of Eqs. (\ref{7}) are unchanged but the
matrix elements of $M_{\alpha \beta }(a,{\bf k})$ are not restricted to
order $a^2$. This new shear rate dependence leads to qualitative differences
from the Navier-Stokes equations (e.g., rheological effects such as shear
thinning, normal stresses). The single parameter of this kinetic model is
the average collision rate for the Boltzmann equation and all dependence on
the interaction potential occurs only through the temperature dependence of
this parameter. We have chosen the simplest case of Maxwell molecules, $%
V(r)\sim r^{-4}$, for which it is a constant. All transport coefficients of
these generalized linear hydrodynamic equations can be calculated exactly as
functions of the shear rate, and the eigenvalues of $M_{\alpha \beta }(a,%
{\bf k})$ can be determined just as in the case of the Navier-Stokes
equations. A long wavelength instability for any value of the shear rate is
found again, now including values of $a$ well outside the limitations of the
Navier-Stokes equation, c.f. Fig.\ref{fig1}.
 We conclude that the instability observed in (\ref{9})
is robust and is not an aberration resulting from the approximations 
(\ref{4}) and (\ref{5}).

The extension of the hydrodynamic equations to larger shear rates using the
BGK model allows comparison with Bird's direct simulation
Monte Carlo method \cite{mareschal,bird}.
There have been significant tests of this method for uniform shear
flow \cite{santos}. The method is so accurate and efficient that
virtually all practical applications of gas kinetic theory far
from equilibrium now use it.
We have used a direct numerical solution to the BGK kinetic equation
to test the stability analysis without 
the intermediate step of constructing a
hydrodynamic description. The solution is constructed as follows. First, the
volume is partitioned into cells and $N$ particles are distributed with
positions and velocities according to a specified initial distribution.
Next, at each finite time step $\tau < $ mean free time, a streaming and
collision stage are computed. The particles are moved in straight lines to
new positions at time $t+\tau $. 
For each particle, the probability of a collision is determined as the
(local) collision frequency times $\tau$.
If a collision
occurs, the velocity is replaced  by a random velocity sampled from the 
local equilibrium distribution.
This collision calculation is performed for all particles and
the whole process is iterated for many time steps.

To confirm our hydrodynamic analysis of the instability based on the BGK
equations an initial state for the hydrodynamic variables is chosen with $%
k=0.1$ and $a=0.5$ (in units of the inverse mean free path and time,
respectively). This corresponds to conditions for which the linear
hydrodynamic equations are unstable. The two unstable modes have complex
eigenvalues so they lead to an oscillatory time dependence with growing
amplitude. The hydrodynamic fields are constructed directly from an
expansion in the eigenvectors of $M_{\alpha \beta }(a,{\bf k})$.  To
illustrate this dynamics we have chosen initial conditions such that $\delta
U_y(0)$ couples only to the unstable modes, while $\delta U_x(0)$ couples to
the both stable and unstable modes. Figure \ref{fig2} shows a comparison of
the results for $\delta U_x(t)$ and $\delta U_y(t)$ as a function of time
with those obtained from a Monte Carlo simulation of the BGK kinetic
equation using the Bird method. The good agreement shows that the
instability is not a consequence of the assumptions behind the hydrodynamic
calculations, and also that these equations provide an accurate description
of the initial stages of the instability. We have performed the simulations
for times an order of magnitude longer than that shown. The maxima and
minima continue to grow, although differences between the theory and
simulation become more significant. This is expected since the linear
hydrodynamic analysis is limited to small amplitudes.

It remains to understand the consequences of this long wavelength
instability. We plan to extend the low density Monte Carlo simulations to
very long times to see if a final stationary state is regained. At high
densities molecular dynamics simulations appear to be stable, except at very
large shear rates where a transition to an ordered state occurs \cite
{erpenbeck}. It is likely that the long wavelength instability considered
here has not been seen due to the finite system sizes considered, i.e. $k >
2\pi /L$ \cite{PL}. We plan to explore longer wavelengths at high densities
using both molecular dynamics and an extension of the Bird method to the
dense fluid Enskog equation \cite{montanero}.

\begin{figure}[t]
\caption{Critical lines for stability determined from 
Eq.~(\protect\ref{7bis}) at $%
n^{*}=0$ (solid curve), $n^{*}=0.2$ (dotted curve), and $n^{*}=0.4$ (dashed
curve). Also shown are the results from the BGK kinetic model for $n^{*}=0$
(dash-dot curve). All are in units of the mean free path and mean free time.}
\label{fig1}
\end{figure}

\begin{figure}[bth]
\caption{Time evolution of (a) $\delta U_{x}(t) $ and (b) $\delta U_{y}(t)$
for $a = 0.5$ and $k = 0.1$ in units of the inverse mean free time and mean
free path.}
\label{fig2}
\end{figure}

\acknowledgments

The research of M.L. and J.W.D. was supported in part by NSF grant PHY
9312723 and the Division of Sponsored Research at the University of Florida.
The research of J.M.M. and A.S. was partially supported by the DGICYT
(Spain) through grants PB94-1021 and PR95-153 and by the Junta de
Extremadura-Fondo Social Europeo.

%\nopagebreak


\begin{references}
\bibitem{edwards}  A. Lees and S. Edwards, J. Phys. C {\bf 5}, 1921 (1972).

\bibitem{dufty1}  J. W. Dufty, J. J. Brey, and A. Santos in {\em Molecular
Dynamics Simulation of Statistical Mechanical Systems}, edited by G.
Ciccotti and W. G. Hoover (North Holland Publ. Co., Amsterdam, 1986), pp.
295--303.

\bibitem{dufty2}  J. W. Dufty, A. Santos, J. J. Brey, and R. Rodr\'{\i}guez,
Phys. Rev. A {\bf 33}, 459 (1986).

\bibitem{BSD}  J. J. Brey, A. Santos, and J. W. Dufty, Phys. Rev. A {\bf 36}%
, 2842 (1987).

\bibitem{hanley}  {\em Nonlinear Fluid Dynamics}, edited by H. Hanley (North
Holland Publ. Co., Amsterdam, 1983).

\bibitem{mareschal}  {\em Microscopic Simulation of Complex Flows}, edited
by M. Mareschal (Plenum Press, New York, 1990).

\bibitem{evans1}  D. J. Evans and A. Baranyai, Phys. Rev. Lett {\bf 67},
2597 (1991); J. J. Brey, A. Santos, and V. Garz\'{o}, Phys. Rev. Lett. {\bf %
70}, 2730 (1993).

\bibitem{evans2}  D. J. Evans, E. G. D. Cohen, and G. P. Morriss, Phys.
 Rev. A {\bf 42}, 5990 (1990).

\bibitem{erpenbeck}  J. Erpenbeck, Phys. Rev. Lett. {\bf 52}, 1333 (1984).

\bibitem{McL}  J.A. McLennan, {\em Introduction to Non-equilibrium
Statistical Mechanics} (Prentice Hall, Englewood Cliffs, NJ, 1989).

\bibitem{hansen} J. P. Hansen and I. McDonald {\em Theory
of Simple Liquids}, 2nd ed. (Academic, New York, 1986)

\bibitem{ferziger}  J. H. Ferziger and H. G. Kaper, {\em Mathematical Theory of
Transport Processes in Gases} (North-Holland, Amsterdam, 1972).

\bibitem{santos}  J. M. Montanero and A. Santos, in {\em Rarefied Gas
Dynamics 19}, edited by J. Harvey and G. Lord (Oxford University Press,
Oxford, 1995); V. Garz\'{o} and A. Santos, Physica A {\bf 213}, 426 (1995);
J. G\'{o}mez
Ord\'{o}\~{n}ez, J. J. Brey, and A. Santos, Phys. Rev. A {\bf 39}, 3038
(1989); {\bf 41}, 810 (1990).

\bibitem{santos2}  A. Santos and J. J. Brey, Physica A {\bf 174}, 355 (1991).

\bibitem{bird}  G. A. Bird. {\em Molecular Gas Dynamics and the Direct
Simulation of Gas Flows} (Clarendon Press, Oxford, 1994).

\bibitem{PL}  J. M. Montanero, A. Santos, and V. Garz\'{o}, Phys. Lett. A 
{\bf 203}, 73 (1995).

\bibitem{montanero}  J. M. Montanero and A. Santos, Phys. Rev. E (
submitted).

\end{references}
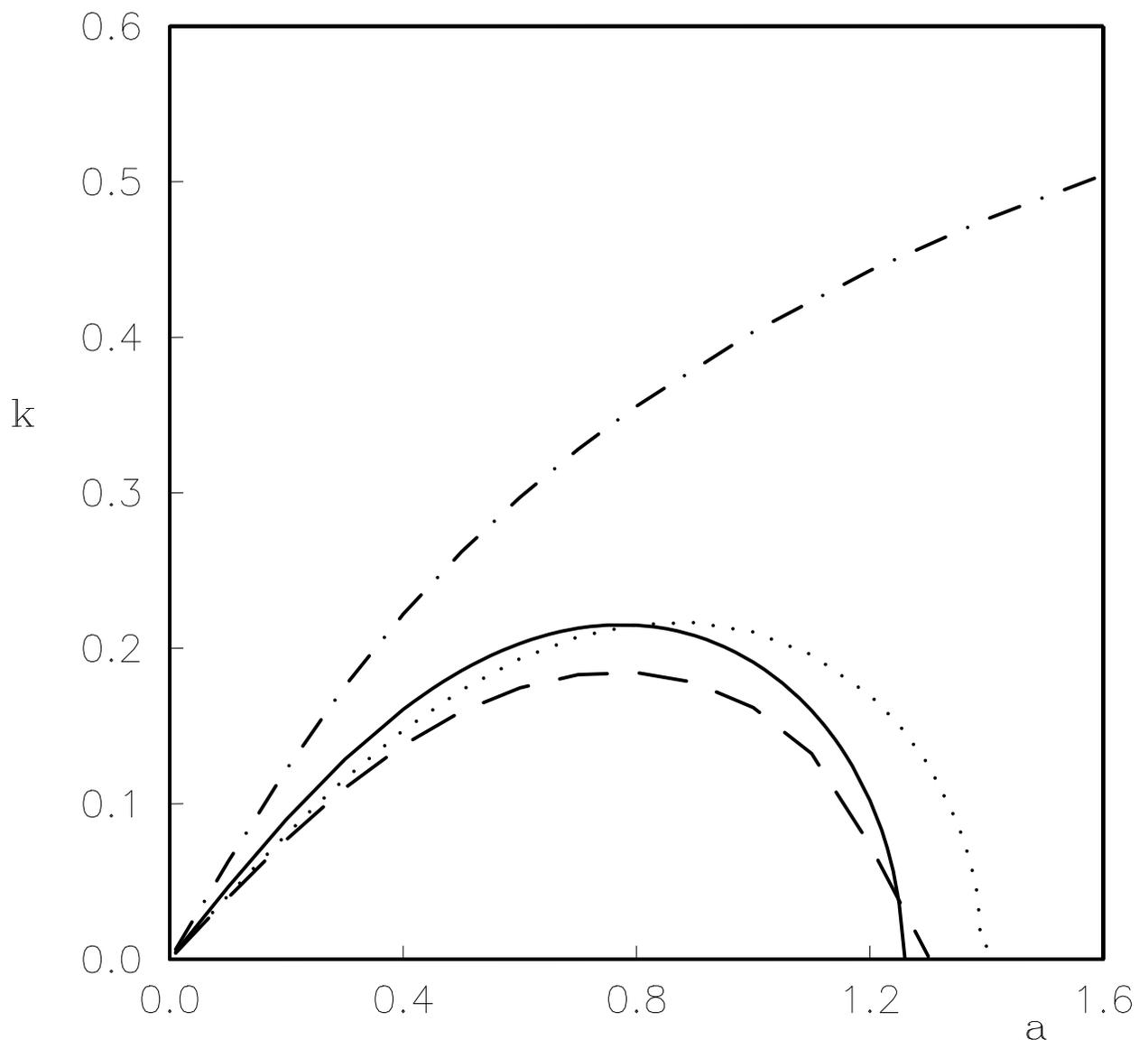
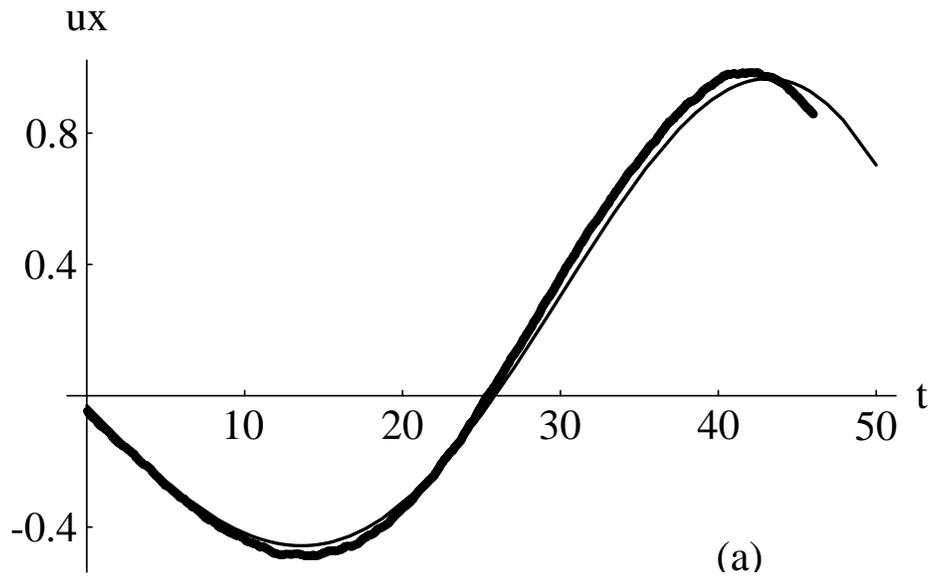
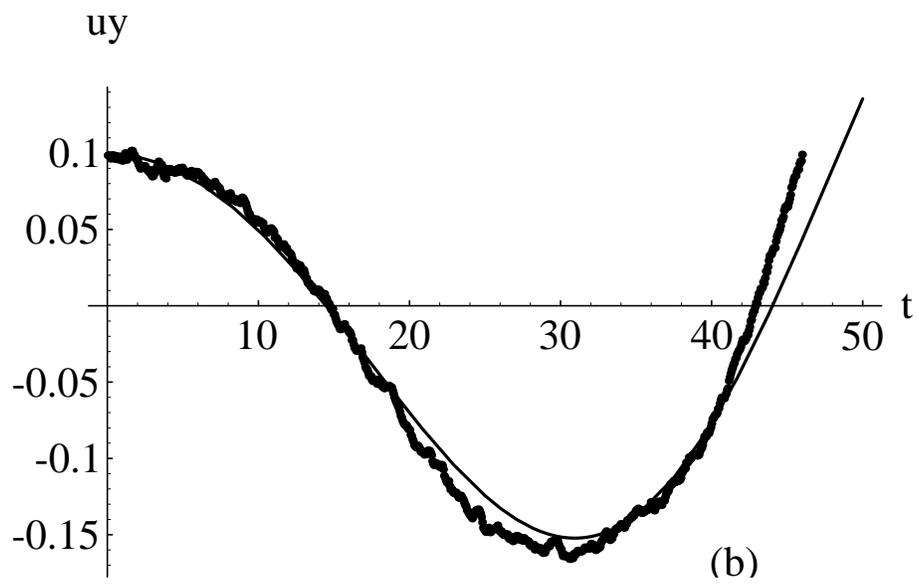
\end{document}